\documentclass{PoS}
\usepackage{citesort}
\newcommand{\eq}{\begin{eqnarray}}
\newcommand{\en}{\end{eqnarray}}

\title{Isospin symmetry breaking}

\ShortTitle{Isospin symmetry breaking}

\author{\speaker{Akaki Rusetsky}\\
        Helmholtz-Institut f\"ur Strahlen- und Kernphysik and
        Bethe Center for Theoretical Physics,\\ University of Bonn,
        53115 Bonn, Germany\\
        E-mail: \email{rusetsky@hiskp.uni-bonn.de}}

\abstract{We
discuss the separation of isospin-symmetric and isospin-breaking contributions
in the hadronic observables within the framework of QCD plus QED. 
Further, we briefly review 
some recent work on the low-energy hadron phenomenology, 
in which the isospin-breaking
effect plays a prominent role.
          }

\FullConference{6th International Workshop on Chiral Dynamics\\
                 July 6-10 2009\\
                 Bern, Switzerland}

\begin{document}

\section{Introduction}

Strong interactions are described by the Lagrangian of pure QCD, whose only
free parameters are the strong coupling constant and the quark masses. 
Most of theoretical predictions, which are made in the framework of pure QCD, 
assume, in addition, isospin symmetry, i.e. the masses of the up 
and down quarks are taken equal. 
In Nature, however, strong interactions do not come alone: the low-energy
processes with the participation of hadrons contain electromagnetic 
effects -- consequently, these processes should be described by the Lagrangian
of QCD plus QED (for simplicity, we neglect weak interactions here).
Therefore, in order to meaningfully compare the
theory with the experiment, one should be able to ``purify'' experimental
data with respect to the isospin-breaking corrections. This task is further
complicated with the fact that the perturbation theory can not be applied to
describe hadronic processes at low energies and one has to resort to the
other techniques, be these the lattice QCD or Chiral 
Perturbation Theory (ChPT).

Specifically, 
one faces the following problem. A physical process is described by
the Lagrangian of QCD plus QED, whose parameters are the strong coupling
constant $g$, the electromagnetic coupling constant $e$ and the (running)
quark masses $m_u,m_d,m_s,\cdots$. We define the {\em isospin-symmetric world}
where $e=0$ and the masses of the up and down quarks are equal. The parameters
of the isospin-symmetric world are the strong coupling constant $\bar g$
and the quark masses $\bar m_u,\bar m_s,\cdots$, with 
$\bar m_u=\bar m_d$ (the bar over a symbol 
denotes the isospin-symmetric world). In general, 
the following questions should be answered:

\begin{itemize}

\item[i)]
How are the parameters $g,e$ and $m_u,m_d,m_s,\cdots$, on the one side, and
$\bar g$ and $\bar m_u,\bar m_s,\cdots$, on the other side, related?

\item[ii)]
How does the separation of isospin-symmetric and isospin-breaking effects
translate to the level of the effective chiral Lagrangian, which is
used to describe hadronic processes at low energy?

\item[iii)]
Once the prescription for separation of the isospin-symmetric and 
isospin-breaking effects is set, what is the systematic procedure of
purifying the hadronic  observables
with respect to the isospin-breaking corrections?

\end{itemize}

Recently, some aspects of the problem
 have been addressed, e.g., in Refs.~\cite{Bijnens,Moussallam,Scimemi,Gegelia}.
Below, we mainly follow the line, adopted
in Ref.~\cite{Scimemi}. Instead of a general reasoning, for illustrative reasons,
we concentrate
on a particular example.

\section{``Purely strong'' pion decay constant}

In this section, we shall study the question, how to extract
the ``purely strong'' pion decay constant $F_\pi$ from experimental
data on the decay of a charged pion with 0 or 1 real photon in the final state.
This problem has a long history,
starting in the pre-ChPT era (see, e.g.~\cite{Sirlin}). 
In ChPT, the decay width of this process has been calculated 
at one~\cite{Talavera} and two~\cite{Cirigliano} loops. Below, we give an
expression of the decay rate at one loop, 
evaluated in ChPT with three flavors~\cite{Talavera}
\eq\label{eq:pil2}
\Gamma(\pi\to\ell\nu_\ell(\gamma))&=&
\frac{G_F^2 |V_{ud}|^2F_0^2m_\ell^2 M_{\pi^+}}{4\pi}\,
\biggl(1-\frac{m_\ell^2}{M_{\pi^+}^2}\biggr)^2
\biggl\{ 1+\frac{8}{F_0^2}\,(L_4^r(M_\pi^2+2M_K^2)+L_5^rM_\pi^2)
\nonumber\\[2mm]
&-&\frac{1}{32\pi^2F_0^2}\,\biggl(2M_{\pi^+}^2\ln\frac{M_{\pi^+}^2}{\mu^2}
+2M_{\pi^0}^2\ln\frac{M_{\pi^0}^2}{\mu^2}+M_{K^+}^2\ln\frac{M_{K^+}^2}{\mu^2}
+M_{K^0}^2\ln\frac{M_{K^0}^2}{\mu^2}\biggr)
\nonumber\\[2mm]
&+&e^2E^r+\frac{e^2}{16\pi^2}\,
\biggl(3\ln\frac{M_\pi^2}{\mu^2}+
H(\frac{m_\ell^2}{M_{\pi^+}^2})\biggr)\biggr\}\, ,
\en
where $m_\ell$ denotes the lepton mass, $M_{\pi^+},M_{\pi^0},M_{K^+},M_{K^0}$
are physical pion and kaon masses, and
$M_\pi^2=2B_0\hat m$, $M_K^2=B_0(m_s+\hat m)$ with
$\hat m=\frac{1}{2}\,(m_u+m_d)$. 
Further, $F_0$ denotes the pion decay constant in the chiral limit,
$B_0$ is proportional to the quark condensate in the chiral limit, $L_i^r$
denote the strong low-energy constants (LECs), $E^r$ is a certain linear combination of the electromagnetic LECs,
the function $H(z)$ stands for the contribution of the photon loops, and
$\mu$ is the scale of the dimensional regularization. Finally, $G_F$ and
 $V_{ud}$ stand for the Fermi constant and the element of the 
Cobayashi-Maskawa matrix.

The reasoning goes as follows. In the isospin-symmetric world,
 the quantity $F_0$ can be related to the charged pion decay constant 
$F_\pi$ through the well-known expression~\cite{GL-npb}
\eq\label{eq:F0}
F_\pi=F_0\biggl\{1+\frac{4}{F_0^2}\,(L_4^r(M_\pi^2+2M_K^2)+L_5^rM_\pi^2)
-\frac{1}{32\pi^2 F_0^2}\,\biggl(2M_\pi^2\ln\frac{M_\pi^2}{\mu^2}
+M_K^2\ln\frac{M_K^2}{\mu^2}\biggr)\biggr\}\, .
\en
Here one has implicitly assumed that passing to the isospin limit amounts 
to setting $e=0$, $m_d=m_u=\hat m$, whereas all other parameters of the theory stay put.
Further, substituting this expression into Eq.~(\ref{eq:pil2}), it is seen that
one may extract the
exact value of $F_\pi$ from the measured value of the decay rate, provided one
makes reliable estimates for the electromagnetic LECs
contained in $E^r$.

In order to demonstrate that this procedure is ambiguous in general, we
calculate the pion decay constant in a model where, unlike QCD, explicit
calculations can be performed. To this end, we invoke the linear $\sigma$-model
to one loop. In the limit when the mass of $\sigma$ becomes much larger
than the pion mass, the low-energy structure of the theory can be described
by ChPT -- with the specific values of the LECs which are determined
by the underlying Lagrangian of the $\sigma$-model.

In the following, we mainly follow  Ref.~\cite{Scimemi}. The Lagrangian
of the linear $\sigma$-model {\em with electromagnetic interactions} is 
given by ${\cal L}={\cal L}_0+{\cal L}_{\sf ct}$, where
\eq\label{eq:L}
{\cal L}_0&=&\frac{1}{2}\,d_\mu \phi^T d^\mu\phi+\frac{m^2}{2}\,\phi^T\phi
-\frac{g}{4}\,(\phi^T\phi)^2
+c\phi^0+\frac{\delta m^2}{2}\,(Q\phi)^T(Q\phi)
\nonumber\\[2mm]
&-&\frac{\delta g}{2}\,(Q\phi)^T(Q\phi)(\phi^T\phi)
-\frac{1}{4}\, F_{\mu\nu}F^{\mu\nu}-\frac{1}{2\xi}\,(\partial_\mu A^\mu)^2\, ,
\en
where $\phi\doteq(\phi^0,\phi^i)$ denotes the 4-component spin-0 field, $A_\mu$ is the electromagnetic field, $F_{\mu\nu}$ is the electromagnetic field tensor,
$\xi$ is the gauge parameter and $Q$ stands for the $4\times 4$ charge matrix, whose non-zero components are $Q_{12}=-Q_{21}=1$. Further, the covariant derivative of the field $\phi$ is defined as $d^\mu\phi=\partial^\mu\phi+(F^\mu+eQA^\mu)\phi$, where
external vector and axial-vector fields are given by
$F_\mu^{0i}=a_\mu^i\, ,\quad F_\mu^{ij}=-\epsilon^{ijk}v_\mu^k$.
Finally, the counterterm Lagrangian
${\cal L}_{\sf ct}=\sum_{a=1}^8\beta_a{\cal O}_a$ includes all operator structures, which are necessary to remove divergences to one loop (for more details, see Ref.~\cite{Scimemi}). The isospin-breaking parameters 
parameters $\delta m^2,\delta g$ are counted at order $e^2$.

The spontaneous chiral symmetry breaking in the model occurs, when $m^2> 0$. 
In this case, the parameter $c$, which describes the explicit breaking  
of chiral symmetry,
determines the mass of the neutral pion (the charged pion mass receives
an additional piece proportional to $e^2$, see~\cite{Scimemi}). The 
mass of the $\sigma$-meson, $M_\sigma$, does not vanish in the chiral limit.

\begin{figure}[t]
\begin{center}
\includegraphics[width=8.cm]{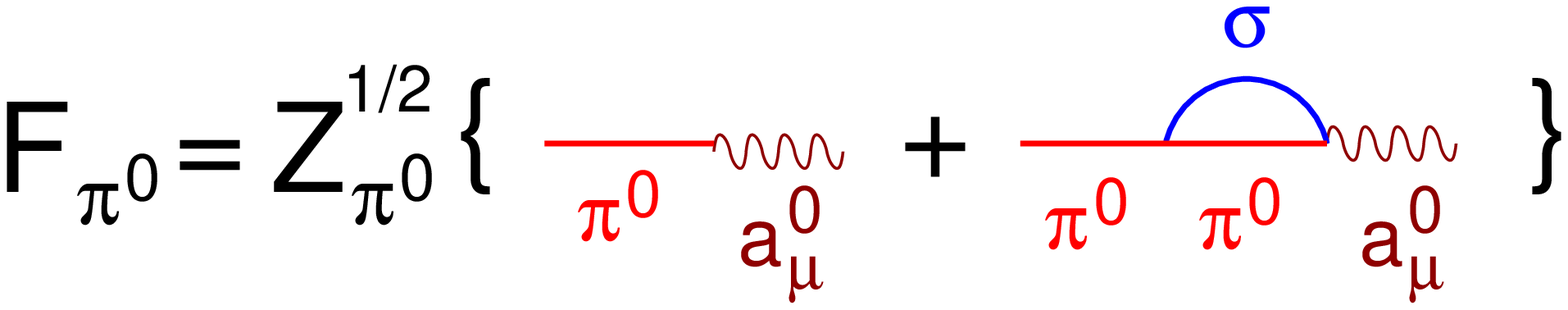}
\end{center}
\caption{The neutral pion decay constant at one loop in the linear $\sigma$-model.}
\label{fig:Fpi0}
\end{figure}

In order to evaluate the pion decay constant, we calculate the matrix
element of an axial current between the vacuum and the neutral pion state
to one loop. This matrix element is determined by the diagrams shown in 
Fig.~\ref{fig:Fpi0}. At the next step, we perform the limit $M_\sigma\to 0$ in
the resulting expressions, and compare the answer with the neutral pion
decay constant, calculated in the two-flavor ChPT~\cite{Knecht}
\eq
F_{\pi^0}=F\biggl\{1-\frac{M_{\pi^+}^2}{16\pi^2F^2}\,\ln\frac{M_{\pi^+}^2}{\mu^2}+\frac{M_{\pi^0}^2}{F^2}\,l_4^r-e^2\sum_i c_ik_i^r\biggr\}\, ,
\en
where $F,l_i^r$ are the $SU(2)$-analogs of $F_0,L_i^r$, and $c_ik_i^r$ stands
for a some linear combination of the electromagnetic LECs. The matching gives
\eq\label{eq:Fki}
F(1-e^2\sum_i c_ik_i^r)=\frac{m}{\sqrt{g}}\,
\biggl\{1-\frac{3g}{16\pi^2}\,\ln\frac{2m^2}{\mu^2}
+\frac{7g}{32\pi^2}\biggr\}\, .
\en
Now, we must move one step further and relate the quantity $F$ to its 
``purely strong'' counterpart, defined in the limit $e=0$. This procedure
is straightforward in Quantum Mechanics but becomes obscure in the framework
of the Quantum Field Theory, where one has to deal with the {\em renormalized}
parameters. The renormalization group (RG) equations, which govern the 
running of these parameters, to one loop are given by (see Ref.~\cite{Scimemi})
\eq\label{eq:RG}
\mu\frac{dm^2}{d\mu}&=&\frac{1}{4\pi^2}\,((3g+\delta g)m^2+g\,\delta m^2)
\\
\mu\frac{dg}{d\mu}&=&\frac{1}{2\pi^2}\,(3g^2+g\,\delta g)
\en

\begin{figure}[t]
\begin{center}

\includegraphics*[width=7.cm, angle=0]{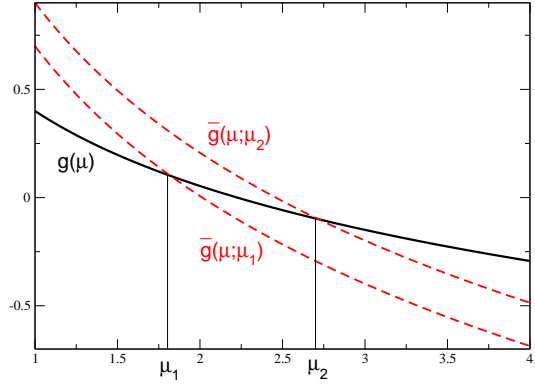}

\end{center}
\caption{Matching of the strong coupling constant in full theory and
in case when the electromagnetic interactions are switched off. The figure
illustrates the uncertainty 
related to the choice of the matching scale.}
\label{fig:gmu}
\end{figure}

Whereas the RG equations in the isospin-symmetric world are obtained
by setting $e=0$ in Eq.~(\ref{eq:RG}) 
\eq\label{eq:RGbar}
\mu\frac{d\bar m^2}{d\mu}&=&\frac{3}{4\pi^2}\,\bar g\bar m^2\, ,
\\
\mu\frac{d\bar g}{d\mu}&=&\frac{3}{2\pi^2}\,\bar g^2\, .
\en
The crucial point here is that the procedure of switching off the 
electromagnetic corrections unambiguously predicts the {\em running} of 
the parameters with respect to the renormalization scale $\mu$, but not the
{\em value} of these parameters at a given $\mu$. The parameters $g(\mu)$,
$m^2(\mu)$ can be, in principle, fixed from the experimental data.
However, this is impossible for $\bar g(\mu)$, $\bar m^2(\mu)$, since
there are no data with electromagnetic interactions switched off. For this,
in order to fix the values of  $\bar g(\mu)$, $\bar m^2(\mu)$, one has
to invoke additional conventions, rendering the definition of the 
isospin-symmetric limit convention-dependent. For example,
one may impose boundary condition on the solutions 
of Eq.~(\ref{eq:RGbar}) at some $\mu=\mu_1$ which will be hereafter referred 
to as the matching scale. Most easily, this can be done by requiring
$g(\mu_1)=\bar g(\mu_1)$ and $m^2(\mu_1)=\bar m^2(\mu_1)$. 
Then, from Eqs.~(\ref{eq:RG}) and (\ref{eq:RGbar}) one obtains
\eq\label{eq:matching}
m^2(\mu)&=&\bar m^2(\mu;\mu_1)+\frac{1}{4\pi^2}\,\ln\frac{\mu}{\mu_1}\,
(\delta g\, m^2+g\,\delta m^2)\, ,
\\
g(\mu)&=&\bar g(\mu;\mu_1)+\frac{1}{2\pi^2}\,\ln\frac{\mu}{\mu_1}\,g\,\delta g\, .
\en
Here, we explicitly indicate the $\mu_1$-dependence of the parameters of the
theory in the isospin symmetry limit. This dependence is illustrated in Fig.~\ref{fig:gmu} where it is shown that, changing the matching scale from $\mu_1$
to $\mu_2$ leads to the change of the value of $\bar g(\mu)$ at a fixed $\mu$.

From Eq.~(\ref{eq:Fki}) one now obtains (remember that the quantity $F$, by
definition, refers to the isospin-symmetric world)
\eq
F=\frac{\bar m}{\sqrt{\bar g}}\,
\biggl\{1-\frac{3\bar g}{16\pi^2}\,\ln\frac{2\bar m^2}{\mu^2}
+\frac{7\bar g}{32\pi^2}\biggr\}\, .
\en
One may directly check that $\mu\frac{d F}{d\mu}=0$, as it should.
However, the quantity $F$ depends on the choice of the matching scale since
$\mu_1\frac{d F}{d\mu_1}\neq 0$. This is the ambiguity inherent to the 
definition of the isospin-symmetry limit. There is no way to avoid this 
ambiguity. On the other hand, a particular combination of $F$ and the
electromagnetic LECs, which is present in Eq.~(\ref{eq:Fki}), is free of this
ambiguity\footnote{This way of reasoning applies also to the quark masses which are present in the Lagrangian of ChPT: these are the parameters of pure QCD and not of QCD plus QED. For instance, $u-$ and $d-$quark masses in the Lagrangian of ChPT have the same RG running, so that their ratio is RG invariant. In analogy to $F$, these masses are also convention-dependent.}.

It turns out that within the framework of the linear $\sigma$-model
it is possible to make a numerical estimate of the size of this uncertainty.
This happens because the quantities $\delta g$ and $\delta m^2$, which
determine the $\mu_1$ dependence of the parameters in Eq.~(\ref{eq:matching}), 
also enter the expression of the charged and neutral pion mass difference.
Using this fact, one obtains the following estimate within the 
linear $\sigma$-model~\cite{Scimemi}
\eq
 F(\mu_1=1~\mbox{GeV})- F(\mu_1=0.5~\mbox{GeV})\simeq 0.1~\mbox{MeV}\, .
\en
That shows that the effect is not purely academic (for example, 
one may compare this
 with the uncertainty
in $F_\pi$, quoted recently in Ref.~\cite{anant}). 

The following conclusions can be drawn:

\begin{itemize}

\item[i)]
The separation of the electromagnetic and strong interactions is not an unambiguous
operation in Quantum Field Theory. We have seen this explicitly on the example
of the linear $\sigma$-model at one loop, 
and we have no reason to believe that this
will be different in QCD.

\item[ii)]
As explained above, this ambiguity can be fixed, e.g., by setting boundary
conditions in the RG equation. As a result, the parameters of the 
isospin-symmetric theory will depend on the choice if the matching scale
$\mu_1$. Alternatively, one may use observable quantities for fixing the
ambiguity. In the example considered in this section, it suffices to set
the pion decay constant and the pion and the $\sigma$ masses in the 
isospin-symmetric world to a given values (known as reference values). This determines the quantities
$\bar g$, $\bar m^2$ and $c$ uniquely. All other observables are expressed
through these parameters.

\item[iii)]
Returning to the issue of extracting the value of $F_\pi$ from the data,
we see that the extraction with an arbitrarily high
precision is not possible, because the quantity $F$ (and hence $F_\pi$
which coincides with $F$ in the chiral limit)
depends on the matching scale $\mu_1$.

\item[iv)]
Using ChPT with virtual photons, which was originally introduced in 
Refs.~\cite{Urech,9506448}, to calculate isospin-breaking
corrections to the ``purely hadronic'' observables implies that the hadronic
LECs stay put in the isospin symmetry limit. In the linear $\sigma$-model
this means that the LECs are expressed through $\bar g$ and $\bar m^2$, so
that they depend on $\mu_1$. On the other hand, the physical quantities 
in the real world do not depend on $\mu_1$. Consequently, the 
$\mu_1$-dependence in the strong LECs must be compensated by a redefinition of
the electromagnetics LECs. In the example with extracting $F_\pi$ from the data
this would mean that there exists an inherent uncertainty which does not allow us to fix the electromagnetic LECs -- and hence $F_\pi$ -- with an arbitrary precision.

\end{itemize}

\section{Isospin-breaking effects in hadronic observables}

From the previous section one may conclude that the definition of the 
(hypothetic) 
isospin-symmetric world, which sets the reference point for the calculation
of the isospin-breaking corrections, is ambiguous. On the other hand,
there exist observable isospin-breaking effects, where the above ambiguity
does not matter. Below, we want to briefly review some recent work where
the isospin-breaking effects play a prominent role, and highlight the issues
which are related to the discussion given in the previous section.

\subsection{Cusps in the $K\to 3\pi$ decays}

Recently, NA48/2 collaboration at CERN observed a pronounced Wigner 
cusp~\cite{wigner} in the invariant mass distribution for the 
three-pion decays of charged kaons~\cite{Batley}.  
Cabibbo in Ref.~\cite{Cabibbo:2004gq} proposed that measuring 
this cusp, which is due to
the presence of the charged and neutral pion mass difference, can be used
for extracting the values of the $\pi\pi$ scattering lengths (originally,
the cusp has been predicted in Ref.~\cite{fonda}, see also~\cite{MMS} where
the cusp in the $\pi\pi$ scattering amplitude is discussed). However,
the accuracy of the one-loop representation of the $K\to 3\pi$ decay 
amplitudes, which is given in Ref.~\cite{Cabibbo:2004gq}, does not suffice
to fit the high-precision experimental data. To this end, in 
Ref.~\cite{Cabibbo:2005ez} one has used analyticity and unitarity
of the $S$-matrix to obtain a theoretical 
representation of the charged and neutral kaon decay amplitudes, valid up to and including
second order in the $\pi\pi$ scattering lengths. The authors of 
Ref.~\cite{Gamiz} have merged the above approach with ChPT, which is only 
used to
calculate the real part of the $K\to 3\pi$ decay amplitude at one loop. 
In Ref.~\cite{CGKR}, the charged pion decays have been 
studied to the same accuracy 
within the non-relativistic effective Lagrangian approach, which
turns out to be a most systematic and convenient tool to address the problem.
In particular, it automatically includes the strictures imposed by
analyticity and unitarity. Furthermore, the inclusion of the electromagnetic effects proceeds 
straightforwardly within this approach~\cite{Krad}. Using two-loop
representation of the decay amplitudes, a precise determination of the
$\pi\pi$ scattering lengths is possible~\cite{Bloch}.
Finally, we mention two more approaches to the same problem.
Ref.~\cite{Zdrahal} uses the dispersive approach. Recently,
there have been attempts~\cite{Tarasov1} to address the problem
in a framework which may be considered as a merger
between $K$-matrix theory and a conventional
quantum-mechanical framework used to study Coulomb interactions in
the final state. In general, this framework
is contradictory and incomplete, since, e.g.,  it fails to reproduce
the correct two-loop structure of the decay amplitude, as well as the full set
of isospin-breaking corrections to it.

It should be pointed out that the non-relativistic approach can be
straightforwardly generalized to study other three-particle decays, like the
decays of the neutral kaons~\cite{KlongLetter}, (for the experiments on
neutral kaons, see Refs.~\cite{Batley,KTeV}), as well as $\eta\to
3\pi$ decays~\cite{Kupsc} and $\eta'\to\eta2\pi$ decays~\cite{Schneider}. From
the recent developments, we also highlight a comprehensive calculation of the
isospin-breaking corrections at order $e^2(m_d-m_u)$ in $\eta\to 3\pi$ decays,
carried out in ChPT to one loop~\cite{Kubis}. These calculations complement
earlier calculations at order $e^2\hat m$, carried out in Ref.~\cite{Wyler}.
For more theoretical and experimental developments on the cusps, see, e.g.,
contributions to this workshop~\cite{Bloch,this}.

We would like to stress that the representation of the decay amplitude,
which is considered in the present subsection, concerns the real world and the observable isospin-breaking effects (like the emerging cusp) only.
No reference is made to the idealized world with no isospin breaking until the
very end, when the relation between the physical $\pi\pi$ scattering amplitudes
in different channels and the $S$-wave $\pi\pi$ scattering lengths $a_0,a_2$
is established (see Ref.~\cite{CGKR} for more detail). Fortunately, 
the isospin-breaking corrections, contained in this relation, at leading order
can be expressed by the charged and neutral pion mass difference and are thus parameter-free~\cite{CGKR}. The electromagnetic LECs, which are the source of ambiguity
discussed in the previous section, appear only at next-to-leading order.
Their contribution are expected to be small and well-controlled.

Finally, we would like to emphasize that
ChPT \cite{weinberg79}, combined with Roy equations,  allows one 
to make very precise predictions
for the values of the S-wave scattering lengths in elastic $\pi\pi$ 
scattering \cite{scattnpb}. For this reason, confronting the experimental results with these
predictions, one may extract important information about the fundamental
properties of QCD at low energy. For instance, the above predictions for the
scattering lengths have been
made in the assumption that the spontaneous chiral symmetry breaking in QCD
proceeds according to the standard scenario (with the large quark condensate).
If the experimentally measured scattering lengths significantly deviate from
the theoretically predicted values, this would indicate that the symmetry
breaking in QCD follows a different path.

\subsection{Isospin-breaking corrections is $K_{e4}$ decays}

$K_{e4}$ decays represent an important source for the determination of the
$\pi\pi$ scattering phase shift and scattering
lengths~\cite{ke4old,ke4NA48/2,Bloch}. The crucial observation which allows
one to link the $K_{e4}$ decay amplitude to the $\pi\pi$ elastic scattering
amplitude is that, according to the Watson theorem, the phases of both
amplitudes coincide. Watson theorem, however, assumes isospin symmetry, which
is violated in Nature. It is therefore justified to ask, how large are the
corrections in the $\pi\pi$ scattering phase, which is extracted from the
$K_{e4}$ decays.

\begin{figure}[t]
\begin{center}
\includegraphics[width=8.cm]{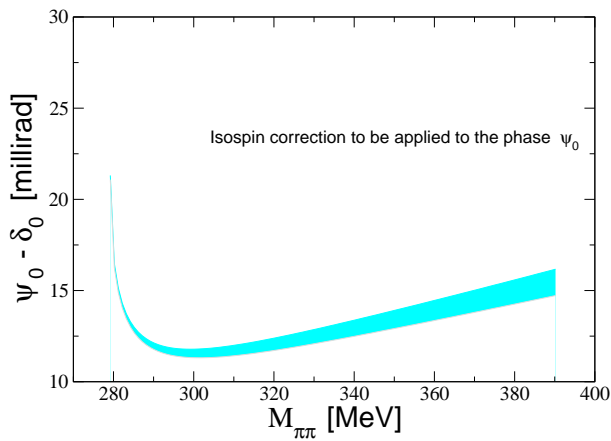}
\end{center}
\caption{The difference between the phase in the $K_{e4}$ amplitude $\psi_0$ and 
the $\pi\pi$ elastic S-wave scattering phase with full isospin $I=0$
in the isospin limit $\delta_0$. The blue band
shows the error caused by the uncertainty in the parameters of the
calculation.  $M_{\pi\pi}$ denotes the invariant mass of the charged pion pair.}
\label{fig:phasediff}
\end{figure}

In fact, in the analysis of the experimental data one has already removed
part of these corrections,
applying the Coulomb factor and using the program PHOTOS \cite{photos} -- 
see Ref.~\cite{ke4NA48/2} for details. This procedure definitely leaves out
some of the isospin-breaking corrections, e.g., those caused by the mass
differences in the isospin multiplets, or direct effects related to the quark
mass difference $m_d-m_u$. Thus, one is faced with an alternative: either one
chooses to deal with the ``raw'' experimental data and removes a full set of
isospin-breaking corrections which can be systematically calculated in ChPT,
or one uses the data already ``corrected'' by PHOTOS and removes additional
isospin-breaking effects, assuming that this can be done separately. Although
the first option is definitely more appealing, it requires a major
calculational effort (see Ref.~\cite{Nehme} for an earlier work in this
direction). Recently, the calculation of the isospin-breaking corrections, using the ``corrected'' data, has
been carried out in Ref.~\cite{ke4} (see also Refs.~\cite{Tarasov2} which
attempt to calculate the effects caused by the pion mass splitting and electromagnetic 
corrections). The findings of Ref.~\cite{ke4} can be summarized as follows:

\begin{itemize}

\item[i)]
The isospin-breaking correction have been calculated to one loop in ChPT.
Virtual photons have been neglected. The charged pion mass is used as the
reference mass for the isospin symmetric world. Both the corrections coming
from the pion mass difference and the effect proportional to $m_d-m_u$, which
originates from the $\pi^0-\eta$ mixing phenomenon, are taken into account. 
Both these effects are of a comparable size.

\item[ii)]
Watson theorem holds only in case of isospin symmetry. When isospin symmetry
is violated, the phases in the $K_{e4}$ decay amplitude and in the $\pi\pi$
elastic amplitude differ. Moreover, they are not related to each other:
for example, the decay amplitude contains some electromagnetic LECs which are
not present in the $\pi\pi$ amplitude.

\item[iii)]
If isospin symmetry is violated, the phase of the $K_{e4}$ decay amplitude has
a {\em cusp below threshold}. The phase no more vanishes at threshold
$s=4M_{\pi^+}^2$.

\item[iv)]
It has been shown that the isospin breaking effect
is quite large (see Fig.~\ref{fig:phasediff}). Moreover, these corrections
strongly affect the values of the S-wave $\pi\pi$ scattering lengths, which
are extracted from the experimental data~\cite{Bloch}. Only after the 
isospin breaking is taken into account, the experimental output for 
the scattering lengths agrees with the theoretical prediction~\cite{scattnpb}.

\end{itemize}

\subsection{Isospin breaking in the $\pi N$ scattering and in 
the pionic hydrogen  observables}

The experiments on hadronic atoms~\cite{DIRAC,PSI,SIDDHARTA}
 provide a beautiful possibility to
directly extract the hadron-hadron scattering lengths from the experimental
data without using the extrapolation to the threshold. However, the accuracy
of the extraction of the scattering lengths critically depends on the
possibility to gain a control on the isospin breaking corrections in the
atomic observables.

In Ref.~\cite{lubovit} the relation of the ground-state energy level 
shift in
the pionic hydrogen and the elastic  $\pi^- p$ threshold scattering amplitude is obtained
at next-to-leading order in isospin breaking. Further, in Ref.~\cite{Zemp}
a similar relation between the width of the ground state and the
charge-exchange $\pi^-p\to \pi^0n$ threshold amplitude has been derived. 
The accuracy of these relations comfortably
matches the existing experimental precision. However, the question of
calculating the isospin-breaking corrections for the threshold amplitudes,
in order to relate these quantities to the S-wave $\pi N$ scattering lengths,
has proven to be more difficult. One invokes ChPT to this end. In
Ref.~\cite{lubovit}, the isospin-breaking corrections to the $\pi^-p$ amplitude
has been obtained at $O(p^2)$, using the charged pion and proton masses as reference
masses in the isospin-symmetric world. In Ref.~\cite{Mojzis} these
calculations have been extended to $O(p^3)$. Finally, in Ref.~\cite{Zemp}
the isospin-breaking corrections to the charge-exchange amplitude have been
evaluated at $o(p^2)$. For the recent review on the subject, see,
e.g.,~\cite{physrep,annual}.

There are, generally, two problems, related to the calculation of these
corrections:

\begin{itemize}

\item[i)]
Using leading order values for the corrections does not always suffice. Loop
corrections can be sizable, see, e.g., Ref.~\cite{Mojzis}.

\item[ii)]
The lowest-order LECs (most notably, the electromagnetic LECs) are poorly
known. Entering the result already at leading order, these LECs are
responsible for the bulk of the uncertainty in the calculations.

\end{itemize}

Although the calculations at higher orders in ChPT can not improve on item
ii), they can definitely check the convergence of the chiral expansion and
lead to a more reliable estimate of the isospin breaking corrections, see item i). This is
the major justification for the comprehensive study of the isospin-breaking
effect in all physical $\pi N$ scattering amplitudes at threshold, which has
been undertaken in Ref.~\cite{Hoferichter} at $O(p^3)$ in the relativistic
baryon ChPT. A careful analysis given in Ref.~\cite{Hoferichter}, which involved both extensive
analytic calculations and a significant numerical effort,  will allow
 one  to carry out the extraction of the $\pi N$
scattering lengths from the pionic hydrogen data in a more controllable manner.

\section{Acknowledgments}

The author would like to thank J\"urg Gasser for numerous interesting 
discussions.
This research is part of the EU HadronPhysics2 project ``Study of
strongly interacting matter'' under the Seventh Framework Programme of EU
(Grant agreement n. 227431). Work supported in part by DFG (SFB/TR
16, ``Subnuclear Structure of Matter''), by the Helmholtz
Association through funds provided to the virtual institute ``Spin
and strong QCD'' (VH-VI-231) and
by the grant of Georgia National Science Foundation
(Grant \# GNSF/ST08/4-401).

\end{document}